\documentclass[lettersize,journal]{IEEEtran}
\usepackage{amsmath,amsfonts}
\usepackage{array}
\usepackage[caption=false,font=normalsize,labelfont=sf,textfont=sf]{subfig}
\usepackage{amsmath} 
\usepackage{textcomp}
\usepackage{multirow} 
\usepackage{stfloats}
\usepackage{booktabs}
\usepackage{url}
\usepackage{verbatim}
\usepackage{graphicx}
\usepackage{cite}
\usepackage{colortbl}
\usepackage{xcolor}
\usepackage{pifont}
\usepackage{multirow}
\newcommand{\cmark}{\ding{51}}%
\newcommand{\xmark}{\ding{55}}%
 \usepackage{placeins} 
\usepackage{mathrsfs}  % 加载 mathrsfs 宏包
\begin{document}

\title{Foundation Model Guided Dual-Branch Co-Adaptation for Source-Free EEG Decoding}

\author{Peiliang Gong*,
        Han Zhang*,
        Zhen Jiang,
        Chenyu Liu,
        Ziyu Jia,
        Xinliang Zhou,
        Daoqiang Zhang,~\IEEEmembership{Senior Member,~IEEE,}
        and Xiaoli Li,~\IEEEmembership{Fellow,~IEEE}
        % <-this % stops a space
\thanks{* Equal contribution.}
\thanks{P. Gong, C. Liu, and X. Zhou are with the College of Computing and Data Science, Nanyang Technological University, 50 Nanyang Avenue, 639798, Singapore}
\thanks{H. Zhang is with the College of Artifical Intelligence and Automation, Hohai University, Changzhou 213022, China}
\thanks{Z. Jiang and D. Zhang are with the College of Artificial
Intelligence, Nanjing University of Aeronautics and Astronautics, Nanjing
211106, China}
\thanks{Z. Jia is with the Brainnetcome Center, Institute of Automation, Chinese Academy of Sciences, Beijing 100190, China}
\thanks{X. Li is with the Information Systems Technology and Design, Singapore University of Technology and
Design, Singapore}
}% <-this % stops a space

\markboth{Journal of \LaTeX\ Class Files,~Vol.~14, No.~8, August~2021}%
{Shell \MakeLowercase{\textit{et al.}}: A Sample Article Using IEEEtran.cls for IEEE Journals}


\maketitle

\begin{abstract}
Source-free domain adaptation (SFDA) provides a practical solution to cross-subject EEG decoding by adapting source-pretrained models to unlabeled target domains without accessing source data. However, existing SFDA methods rely solely on the limited internal knowledge of source-pretrained models, leading to inferior cross-domain generalization and unreliable pseudo-labels. Although EEG Foundation Models (FMs) pretrained on large-scale data exhibit strong generalizability, their potential in SFDA remains largely unexplored.
To this end, we propose \textbf{FUSED}, a \textbf{\underline{F}}oundation-g\textbf{\underline{U}}ided \textbf{\underline{S}}ource-free \textbf{\underline{E}}EG \textbf{\underline{D}}ecoding framework that integrates a large-scale FM with a compact Specialist Model (SM) via dual-branch co-adaptation. 
Specifically, we introduce a Co-adaptation mechanism equipping both branches with linear and prototype views, enabling cross-branch pseudo-label generation. Additionally, we design a Consensus Filtering Mechanism that exploits the FM's inherent stability to identify high-quality samples, along with a Two-Stage Pseudo-Label Refinement scheme to suppress error accumulation through cross-branch arbitration. Finally, we calibrate the FM's decision boundaries via mutual information maximization with the SM, followed by knowledge distillation from FM to SM, forming a principled calibrate-then-distill pipeline.
To our knowledge, FUSED is the first work to leverage EEG FMs within the SFDA framework for cross-subject EEG decoding. Extensive experiments across three EEG paradigms, including motor imagery, emotion recognition, and SSVEP, demonstrate consistent state-of-the-art performance, validating the effectiveness of foundation-guided synergy for robust and privacy-preserving EEG decoding.
\end{abstract}

\begin{IEEEkeywords}
    EEG, Brain-computer Interface, Source-free Domain Adaptation, Foundation Model
\end{IEEEkeywords}
\maketitle

\section{Introduction}
%%%%
% Introduction narrative:
% Paragraph 1: Background
% ├── EEG: non-invasive BCI modality, broad applications
% ├── Deep learning: end-to-end paradigm success
% └── Challenge: non-stationarity + inter-subject variability

% Paragraph 2: From UDA to SFDA
% ├── UDA: addresses domain shift, but requires source data access
% ├── SFDA: practical alternative without source data
% ├── SFDA success in vision tasks
% └── SFDA challenge in EEG: single-model self-training → error accumulation

% Paragraph 3: Foundation Models
% ├── FM emergence: large-scale pre-training, strong generalization
% ├── FM-guided SFDA success in vision tasks
% ├── Challenge: EEG FM ≠ Vision-Language FM (no zero-shot)
% ├── Our findings [Figure 1(a)]: FM has better feature space + stable confidence
% ├── But: FM doesn't consistently beat SM on specific tasks
% ├── Trade-off: FM (generalization) vs. SM (task-specific)
% ├── Insight [Figure 1(b)]: FM-guided paradigm vs. single-model
% ├── Evidence [Figure 1(c)]: FM-guided yields substantial gains
% └── Research question: How to synergistically combine FM and SM?

% Paragraph 4: Our Method (FUSED)

% Paragraph 5: Contributions
%%%%

Electroencephalography (EEG) is a non-invasive neural recording technique that has been widely adopted in Brain-Computer Interface (BCI) applications, including motor imagery, emotion recognition, and Steady-State Visual Evoked Potential (SSVEP)~\cite{zheng2015investigating, altaheri2023deep, li2024facilitating, gong2025tahag}. With the advent of deep learning (DL), end-to-end paradigms that learn directly from raw EEG signals have achieved remarkable success, bypassing the need for laborious manual feature engineering~\cite{chen2022toward}. However, EEG signals are inherently \textit{non-stationary} and exhibit pronounced \textit{inter-subject variability}, causing pre-trained classifiers to suffer severe performance degradation when deployed across different individuals.

To address such domain shifts, Unsupervised Domain Adaptation (UDA) methods have been developed to align feature distributions between source and target subjects~\cite{jimenez2020custom, zhao2021label, shen2025ua}. However, UDA requires concurrent access to both source and target data during adaptation—an assumption often violated in real-world BCI deployments due to privacy regulations or limited on-device storage~\cite{xia2022privacy}. Source-Free Domain Adaptation (SFDA) has emerged as a practical alternative, enabling model calibration on target domains without accessing original source data. While SFDA has achieved considerable success in computer vision~\cite{liang2020we, li2024comprehensive, liu2021source, yang2022attracting}, its application to EEG decoding remains challenging. Existing EEG-based SFDA methods rely on single-model self-training with pseudo-labels~\cite{zhao2023source, DBLP:journals/corr/abs-2504-03707}, which is particularly vulnerable to the stochastic dynamics of neural signals, leading to noisy pseudo-labels and error accumulation.

\begin{figure}[tbp]
    \centering
    \includegraphics[width=1.0\columnwidth]{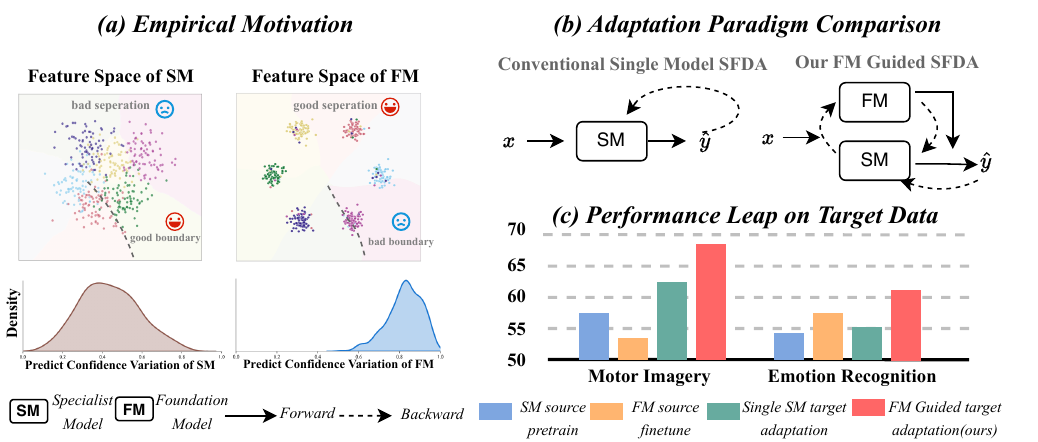}
    \caption{\textbf{Illustration of FUSED: Motivation, Paradigm, and Results.} 
    (a) FM provides more discriminative feature spaces and reliable confidence scores than SM. 
    (b) Architecture of the proposed FM-guided SFDA vs. conventional single-model SFDA. 
    (c) Performance comparison on target domain for Motor Imagery and Emotion Recognition, showing the significant gain achieved by our new paradigm.}
    \label{fig:teaser}
\end{figure}

Recently, Foundation Models (FMs) pre-trained on large-scale data have demonstrated strong generalization capabilities across various domains. In computer vision (CV), FM-guided SFDA methods~\cite{tang2024source, DBLP:conf/iclr/0001SG0ZZ25} have shown promising results by leveraging vision-language models as external knowledge anchors. This success raises a natural question: \textit{Can EEG Foundation Models similarly guide SFDA for cross-subject EEG decoding?} To investigate this, we conduct empirical analyses comparing FMs and compact Specialist Models (SMs). As shown in Figure~\ref{fig:teaser}(a), FMs pre-trained on large-scale heterogeneous EEG corpora exhibit more discriminative feature spaces with clearer class boundaries. Moreover, the FM's prediction confidence distribution is notably more stable, suggesting stronger robustness against domain shift. These observations indicate that FMs could serve as reliable guides for pseudo-label generation in SFDA. However, unlike vision-language models with zero-shot capabilities, EEG FMs require task-specific fine-tuning~\cite{DBLP:conf/icml/LeeBPALZ25}. Furthermore, recent benchmarks~\cite{liu2026eeg, yang2026eeg} reveal a counterintuitive finding: despite their superior generalization, fine-tuned FMs do not consistently outperform compact SMs on specific EEG tasks, likely due to the misalignment between generic pre-training objectives and fine-grained task requirements. These findings suggest a fundamental trade-off: FMs offer robust generalization but may lack task-specific precision, whereas SMs capture discriminative nuances but are susceptible to label drift under domain shift. Neither model alone is sufficient—yet their complementary strengths suggest a promising synergy. As illustrated in Figure~\ref{fig:teaser}(b), unlike conventional single-model SFDA that relies solely on SM's limited internal knowledge, an FM-guided paradigm enables cross-branch knowledge exchange. Figure~\ref{fig:teaser}(c) further validates this insight: FM-guided adaptation yields substantial performance gains over both SM source pre-training and standalone FM fine-tuning on motor imagery and emotion recognition tasks. These observations motivate our research question: \textit{\textbf{How can we synergistically integrate the robust generalization of FMs with the localized discriminative power of SMs to achieve stable and precise adaptation under source-free constraints?}}

To address this challenge, we propose \textbf{FUSED}, a \textbf{\underline{F}}oundation-g\textbf{\underline{U}}ided \textbf{\underline{S}}ource-free \textbf{\underline{E}}EG \textbf{\underline{D}}ecoding framework that bridges universal neural heuristics with task-specific decoding through dual-branch co-adaptation. Specifically, we equip both FM and SM branches with dual predictive views—linear classification and prototypical proximity—enabling cross-branch consensus for robust prediction. To ensure pseudo-label quality, we introduce a Consensus Filtering Mechanism that exploits the FM's inherent stability by requiring agreement between its linear and prototypical predictions. We further develop a Two-Stage Pseudo-Label Refinement scheme: it first verifies cross-branch linear agreement, and upon disagreement, invokes structural arbitration by comparing relative prototypical distances to determine the most reliable label. Finally, we implement a \textit{calibrate-then-distill} pipeline that first aligns the FM's decision boundaries via mutual information maximization, then transfers these refined heuristics to the SM through knowledge distillation.

The main contributions of this paper can be summarized as follows:
\begin{itemize}
    % \item We present FUSED, the first framework to leverage large-scale EEG Foundation Models within the SFDA paradigm, providing a robust and privacy-preserving solution for cross-subject EEG decoding.
    
    \item To our knowledge, we are the first to leverage large-scale EEG Foundation Models within the SFDA paradigm. The proposed FUSED framework provides a robust and privacy-preserving solution for cross-subject EEG decoding.
    
    \item We introduce a Dual-View Co-adaptation mechanism with Consensus Filtering and Two-Stage Pseudo-Label Refinement, which exploits cross-branch structural arbitration to effectively suppress error accumulation during target-domain self-training.
    
    \item Extensive experiments across three EEG paradigms—motor imagery, emotion recognition, and SSVEP—demonstrate consistent state-of-the-art performance, validating the efficacy of foundation-guided synergy.
\end{itemize}

\section{Related Work}

\subsection{Source-Free Domain Adaptation}
Existing SFDA methods can be broadly categorized into data-based and model-based approaches. Data-based methods convert SFDA to traditional UDA by generating source-like samples via generative adversarial networks~\cite{ijcai2021p402,9530705} or selecting target subsets that resemble the source distribution~\cite{DBLP:journals/ml/DuYCLJXW24}. Model-based methods fine-tune source pre-trained models on the target domain through various strategies, including pseudo-labeling~\cite{9528982, DING202392, liang2020we}, entropy minimization~\cite{mao2024source, ahmed2021unsupervised}, auxiliary tasks~\cite{gong2025temporal, ragab2025evidentially}, and contrastive learning~\cite{huang2021model, zhang2022divide}. However, these approaches rely solely on the internal knowledge and limited decision boundaries of source pre-trained models, constraining their adaptation capacity.
Recently, foundation models have been incorporated to guide SFDA in computer vision~\cite{DBLP:conf/iclr/0001SG0ZZ25, yao2026beyond, lee2025duet, tang2024source}. ProDe~\cite{DBLP:conf/iclr/0001SG0ZZ25} introduces proxy denoising to iteratively refine predictions, while DUET~\cite{lee2025duet} leverages consensus between target and foundation models for pseudo-labeling. However, these methods are tailored for vision-language models with strong zero-shot capabilities. In contrast, EEG FMs, despite exhibiting robust cross-task generalization, often fail to consistently outperform compact SMs on specific EEG tasks~\cite{liu2026eeg, yang2026eeg}, likely due to misalignment between generic pre-training objectives and fine-grained task requirements. Our approach bridges this gap through a dual-branch co-adaptation mechanism that synergistically combines the strengths of both model types.

\begin{figure*}[t]
    \centering
    % 裁剪右侧 55mm 的空白
    % \includegraphics[trim=0mm 0mm 55mm 0mm, clip, width=0.97 \textwidth]{framework.pdf} 
    \includegraphics[width=0.97\textwidth]{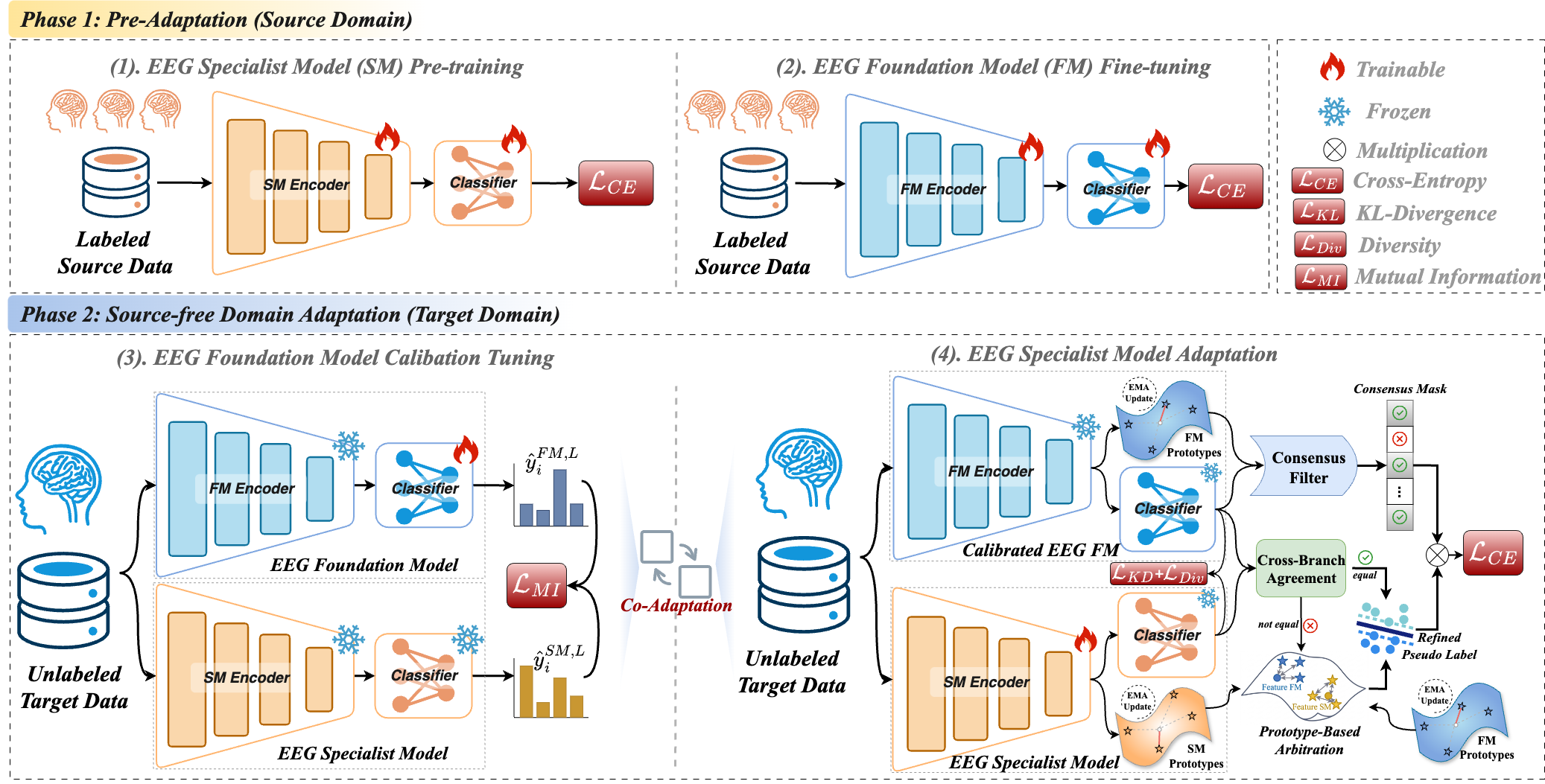}
    \caption{Overview of the FUSED framework. Phase 1 (Pre-Adaptation): The SM is pre-trained and the FM is fine-tuned on labeled source data using cross-entropy supervision. Phase 2 (Source-Free Adaptation): Following a calibrate-then-distill paradigm, (3) the FM classifier is first calibrated via mutual information maximization ($\mathcal{L}_{\text{MI}}$) to align with the target distribution, then (4) the calibrated FM guides SM adaptation through consensus-filtered pseudo-labeling ($\mathcal{L}_{\text{CE}}$), knowledge distillation ($\mathcal{L}_{\text{KD}}$), and diversity regularization ($\mathcal{L}_{\text{Div}}$). The Consensus Filter identifies high-quality samples via FM dual-view agreement, while the Two-Stage Pseudo-Label Refinement resolves cross-branch disagreements through prototype-based arbitration. Fire and snowflake icons indicate trainable and frozen components, respectively.}
    \label{fig:framework}
\end{figure*}

\subsection{EEG-based Decoding}
EEG decoding has evolved from conventional machine learning to end-to-end DL approaches. Early methods relied on handcrafted features such as differential entropy and power spectral density, followed by shallow classifiers~\cite{lan2016real, zheng2015investigating}. With the advent of DL, diverse architectures have been developed to capture the complex spatiotemporal dynamics of neural signals: CNN-based models for local spatial patterns~\cite{lawhern2018eegnet, ding2022tsception}, Graph Convolutional Networks for functional connectivity~\cite{song2018eeg}, RNN-based approaches for temporal dependencies~\cite{supakar2022deep}, and Transformers for long-range interactions~\cite{song2022eeg, ding2025emt}. Despite their success in subject-dependent settings, these models struggle with domain shifts induced by inter-subject variability.
To address cross-subject generalization, various adaptation techniques have been explored. DResNet~\cite{ma2019reducing} employs adversarial training to extract domain-invariant features, CLISA~\cite{shen2022contrastive} leverages contrastive learning to maximize cross-subject representation similarity, and BrainUICL~\cite{zhou2025brainuicl} introduces continual learning for incremental subject-specific adaptation. However, these task-specific models lack universal generalization capability.
Inspired by large-scale pre-training in NLP and CV, EEG FMs~\cite{DBLP:conf/iclr/WangZLZJLL025, DBLP:conf/iclr/JiangZL24, yang2023biot, ma2025codebrain} have emerged as a promising direction. Through self-supervised pre-training on massive heterogeneous datasets, these models exhibit strong cross-task generalization. The prevailing paradigm involves fine-tuning FMs on labeled source data and directly evaluating on target domains. However, recent studies~\cite{liu2026eeg, yang2026eeg} reveal that fine-tuned FMs do not consistently outperform compact SMs on specific EEG tasks, likely due to the misalignment between generic pre-training objectives and fine-grained task requirements.
These observations motivate our work: rather than relying on either FM or SM alone, we propose to synergistically leverage both within a source-free adaptation framework, exploiting the FM's robust generalization to guide the SM's task-specific adaptation without accessing source data.

\section{Methodology}
\subsection{Problem Formulation}
We consider a labeled source domain $\mathcal{D}_s = \{(\mathbf{X}_s^i, y_s^i)\}_{i=1}^{n_s}$ and an unlabeled target domain $\mathcal{D}_t = \{\mathbf{X}_t^j\}_{j=1}^{n_t}$, where $\mathbf{X} \in \mathbb{R}^{C \times T}$ denotes an EEG sample with $C$ channels and $T$ time points, and $y \in \{1, \ldots, K\}$ is the class label among $K$ categories. Following the closed-set assumption, source and target domains share an identical label space. Under the SFDA setting, the target domain contains no labels, and source data is strictly inaccessible during adaptation.
\subsection{Overview}
% \paragraph{Framework Overview.}
As illustrated in Figure~\ref{fig:framework}, FUSED comprises two parallel branches: a FM employing large-scale pre-trained architectures for robust generalization, and a SM adopting a compact task-specific architecture for discriminative precision. Both branches share a unified encoder-classifier structure: input $\mathbf{X}$ is processed by backbone $\mathcal{F}$ to extract features $\mathbf{z} \in \mathbb{R}^{D}$, which are then mapped by classifier $\mathcal{G}$ to prediction $\mathbf{p} = \mathrm{softmax}(\mathcal{G}(\mathbf{z}))$. Additionally, each branch maintains class prototypes $\{\mathbf{c}_k\}_{k=1}^K$ updated via exponential moving average (EMA), enabling a dual-view prediction scheme through both linear classification and prototype-based similarity.

% \paragraph{Two-Phase Pipeline.}
FUSED operates in two phases. In \textit{Phase 1: Pre-Adaptation}, the SM is pre-trained and the FM is fine-tuned on labeled source data $\mathcal{D}_s$ using standard cross-entropy supervision, establishing task-specific decision boundaries for both branches. In \textit{Phase 2: Source-Free Adaptation}, we adopt a \textit{calibrate-then-distill} paradigm on unlabeled target data $\mathcal{D}_t$: the FM classifier is first calibrated via mutual information maximization to align with the target distribution, after which the calibrated FM guides SM adaptation through knowledge distillation and consensus-filtered pseudo-labeling. Throughout adaptation, FM backbone and the SM classifier remain frozen to preserve source-learned representations, with only the FM classifier and SM backbone being updated. We elaborate each component in the following sections.

\subsection{Dual-View Representation}
\label{sec:dual_view}
To generate robust predictions under domain shift, we equip each branch with two complementary views: a \textit{linear view} and a \textit{prototype view}. The linear view performs classification via the learned classifier $\mathcal{G}$, offering flexibility to adapt to new distributions. The prototype view classifies samples based on feature-prototype similarity, providing stability through class-center representations that are less sensitive to individual noisy samples.
\paragraph{Linear View.}
The linear view directly applies the classifier to extracted features. For branch $b \in \{\text{FM}, \text{SM}\}$, the predicted probability over $K$ classes is:
\begin{equation}
\mathbf{p}^{b,\text{L}} = \mathrm{softmax}(\mathcal{G}_b(\mathbf{z}^b))
\label{eq:linear_view}
\end{equation}
where the superscript L denotes the linear view. The predicted label is $\hat{y}^{b,\text{L}} = \arg\max_k p_k^{b,\text{L}}$.

\paragraph{Prototype View.}
Each branch maintains a set of class prototypes $\{\mathbf{c}_k^b\}_{k=1}^K$ representing class centers in the feature space. We initialize prototypes using the $\ell_2$-normalized classifier weights, ensuring alignment between the two views at the start of adaptation.

During adaptation, prototypes are refined via EMA using high-confidence samples. To assess confidence, we compute the prediction margin—the difference between the top two probabilities:
\begin{equation}
\text{margin}_i^{b} = p_{i,\text{top1}}^{b,\text{L}} - p_{i,\text{top2}}^{b,\text{L}}
\label{eq:margin}
\end{equation}
Only samples exceeding a threshold $\eta$ contribute to prototype updates:
\begin{equation}
\mathbf{c}_k^b \leftarrow \mu \mathbf{c}_k^b + (1-\mu) \frac{\sum_{i=1}^{N} \bar{\mathbf{z}}_i^b \cdot \mathbb{I}[\hat{y}_i = k \land \text{margin}_i^{b} > \eta]}{\sum_{i=1}^{N} \mathbb{I}[\hat{y}_i = k \land \text{margin}_i^{b} > \eta]}
\label{eq:prototype_update}
\end{equation}
where $\mu$ is the momentum coefficient, $\bar{\mathbf{z}}_i^b$ denotes the $\ell_2$-normalized feature vector, and $\mathbb{I}[\cdot]$ is the indicator function.

The prototype view computes class probabilities via scaled cosine similarity:
\begin{equation}
p^{b,\text{P}}_{i,k} = \frac{\exp(\tau \cdot \mathrm{sim}(\bar{\mathbf{z}}_i^b, \mathbf{c}_k^b))}{\sum_{j=1}^{K} \exp(\tau \cdot \mathrm{sim}(\bar{\mathbf{z}}_i^b, \mathbf{c}_j^b))}
\label{eq:proto_pred}
\end{equation}
where $\tau$ is a temperature parameter, $\mathrm{sim}(\cdot, \cdot)$ denotes cosine similarity, and superscript P denotes the prototype view. The predicted label is $\hat{y}^{b,\text{P}} = \arg\max_k p_k^{b,\text{P}}$.

\subsection{Foundation Model Calibration}
\label{sec:fm_calibration}
Although the FM provides robust representations through large-scale pre-training, its classifier—trained on source data—may exhibit misaligned decision boundaries for the target distribution. Directly using such a classifier as a teacher would propagate source-biased supervision to the SM. To address this, we calibrate the FM classifier by maximizing MI between FM and SM predictions, encouraging the FM to respect the target manifold structure captured by the SM while preserving its generalization capability.

Let $\mathbf{P} \in \mathbb{R}^{K \times K}$ denote the joint probability distribution over the target domain, where entry $P_{jk}$ represents the probability of the FM predicting class $j$ and the SM predicting class $k$:
\begin{equation}
P_{jk} = \frac{1}{|\mathcal{D}_t|} \sum_{i \in \mathcal{D}_t} p_{i,j}^{\text{FM},\text{L}} \cdot p_{i,k}^{\text{SM},\text{L}}
\label{eq:joint_prob}
\end{equation}
The marginal distributions are $\bar{P}_j^{\text{FM}} = \sum_{k} P_{jk}$ and $\bar{P}_k^{\text{SM}} = \sum_{j} P_{jk}$. The MI calibration loss is then defined as:
\begin{equation}
\mathcal{L}_{\text{MI}} = -\sum_{j=1}^{K} \sum_{k=1}^{K} P_{jk} \left[\log P_{jk} - \log \bar{P}_j^{\text{FM}} - \log \bar{P}_k^{\text{SM}}\right]
\label{eq:mi_loss}
\end{equation}
Minimizing $\mathcal{L}_{\text{MI}}$ maximizes the mutual information between the two branches, which simultaneously encourages prediction consensus and maintains class diversity.

During this phase, only the FM classifier $\mathcal{G}_{\text{FM}}$ is updated while the FM backbone $\mathcal{F}_{\text{FM}}$ remains frozen, preserving the pre-trained representations. This calibration produces a reliable teacher signal for subsequent knowledge distillation to the SM.

\subsection{Specialist Model Adaptation}
\label{sec:sm_adaptation}
With the calibrated FM serving as a reliable teacher, we now adapt the SM to the target domain. This involves three key components: a consensus filtering mechanism to identify high-quality samples, a two-stage pseudo-label refinement scheme to generate reliable supervision, and a joint training objective combining pseudo-labeling, knowledge distillation, and diversity regularization.

\subsubsection{Consensus Filtering Mechanism}
Single-model self-training is prone to confirmation bias, where prediction errors accumulate through noisy pseudo-labels. To mitigate this, we exploit the FM's inherent stability from large-scale pre-training to identify reliable samples. Specifically, we define a consensus mask $m_i \in \{0, 1\}$ that selects samples where the FM's dual views agree:
\begin{equation}
m_i = \mathbb{I}[\hat{y}_i^{\text{FM},\text{P}} = \hat{y}_i^{\text{FM},\text{L}}]
\label{eq:consensus_mask}
\end{equation}
where $\hat{y}_i^{\text{FM},\text{P}}$ and $\hat{y}_i^{\text{FM},\text{L}}$ denote the FM's prototype-view and linear-view predictions, respectively. A sample is deemed high-quality ($m_i = 1$) only when both views yield the same prediction, indicating consistent classification across complementary perspectives. Consensus-filtered samples are used for pseudo-label supervision.

\subsubsection{Two-Stage Pseudo-Label Refinement}
While the FM provides robust generalization, the SM better captures task-specific nuances. We leverage both branches through a two-stage refinement process that combines cross-branch consensus with structural arbitration.

\paragraph{Stage 1: Cross-Branch Agreement.}
For each target sample, we first examine whether the linear views of both branches agree. If $\hat{y}_i^{\text{FM},\text{L}} = \hat{y}_i^{\text{SM},\text{L}}$, we adopt this consensus prediction as the refined pseudo-label, as agreement between two heterogeneous models indicates high reliability.

\paragraph{Stage 2: Prototype-Based Arbitration.}
Upon disagreement, we invoke structural arbitration by evaluating feature-prototype similarities across both branches. The refined label is assigned to the class with the highest similarity in either branch:
\begin{equation}
\hat{y}_{i} = 
\begin{cases} 
\hat{y}_i^{\text{FM},\text{L}}, & \text{if } \hat{y}_i^{\text{FM},\text{L}} = \hat{y}_i^{\text{SM},\text{L}} \\[6pt]
\arg\max_k \max\{\mathrm{sim}(\bar{\mathbf{z}}_{i}^{\text{FM}}, \mathbf{c}_k^{\text{FM}}), \mathrm{sim}(\bar{\mathbf{z}}_{i}^{\text{SM}}, \mathbf{c}_k^{\text{SM}})\}, & \text{otherwise}
\end{cases}
\label{eq:pseudo_label}
\end{equation}
This hierarchical strategy ensures that pseudo-labels are determined by the most confident prediction across branches, effectively suppressing error accumulation.

% 对比表格1
\begin{table*}[htbp]
\centering
\caption{Cross-subject classification accuracy (\%) of different methods on the BCIC-IV-2a using the Leave-One-Subject-Out (LOSO) protocol. The best results are highlighted in \textbf{bold}. And the second best results are \underline{underlined}}
\label{tab:loso_results}
% \small 
\renewcommand{\arraystretch}{1.1} % 默认是 1，调小可以压缩行高
\setlength{\tabcolsep}{8.5pt} 
\begin{tabular}{lcccccccccc}
\toprule
\textbf{Method} & \textbf{S1} & \textbf{S2} & \textbf{S3} & \textbf{S4} & \textbf{S5} & \textbf{S6} & \textbf{S7} & \textbf{S8} & \textbf{S9} & \textbf{Avg.} \\ 
\midrule
Source-only & 67.19 & 48.26 & 73.65 & 48.96 & 48.61 & 52.08 & 67.01 & 68.23 & 64.24 & 57.80 \\
\midrule
SHOT\cite{liang2020we}    & 72.40 & 49.48 & \underline{86.28} & 53.30 & \underline{53.30} & 58.16 & 76.56 & 75.66 & \underline{75.03} & 66.68 \\
AaD\cite{yang2022attracting}    & 69.36 & \underline{54.86} & 83.36 & 53.29 & 52.85 & 60.15 & 75.05 & 71.02 & 72.72 & 65.85 \\
NRC\cite{yang2021exploiting}    & 69.31 & 53.47 & 84.14 & 53.94 & 53.50 & 60.06 & 69.31 & 72.54 & 72.77 & 65.44 \\
MAPU\cite{ragab2023source}    & \textbf{75.52} & 52.95 & 82.12 & 55.21 & 48.49 & 58.16 & 76.39 & 73.26 & 68.06 & 65.57 \\
$\text{SF(DA)}^2$\cite{DBLP:conf/iclr/Hwang0SY24}    & \underline{74.48} & 53.12 & \textbf{86.98} & \textbf{60.42} & 46.35 & \underline{60.37} & 76.91 & \textbf{76.56} & 74.13 & \underline{67.70} \\
E-MAPU\cite{ragab2025evidentially}    & 72.52 & 52.92 & 83.17 & 55.42 & 49.57 & 59.60 & \underline{77.12} & 72.06 & 74.33 & 66.27 \\
\midrule
\rowcolor{gray!20}
\textbf{FUSED (Ours)} & 72.57 & \textbf{55.90} & \textbf{86.98} & \underline{56.65} & \textbf{54.86} & \textbf{62.85} & \textbf{77.60} & \underline{76.39} & \textbf{75.17} & \textbf{68.77} \\
\bottomrule
\end{tabular}
\end{table*}

\subsubsection{Training Objectives}

The SM backbone $\mathcal{F}_{\text{SM}}$ is optimized through three complementary objectives, while the SM classifier $\mathcal{G}_{\text{SM}}$ remains frozen to preserve source-domain decision boundaries.

\paragraph{Masked Cross-Entropy Loss.}
We supervise the SM using refined pseudo-labels, weighted by the consensus mask to ensure only reliable samples contribute:
\begin{equation}
\mathcal{L}_{\text{CE}} = -\frac{1}{\sum_{i} m_i} \sum_{i \in \mathcal{D}_t} m_i \cdot \sum_{k=1}^{K} \mathbb{I}(\hat{y}_{i} = k) \log p_{i,k}^{\text{SM},\text{L}}
\label{eq:ce_loss}
\end{equation}
where $m_i$ is the consensus mask from Eq.~\eqref{eq:consensus_mask} and $\hat{y}_i$ is the refined pseudo-label from Eq.~\eqref{eq:pseudo_label}.

\paragraph{Knowledge Distillation Loss.}
To transfer the calibrated FM's knowledge, we align the SM's predictions with the FM's soft outputs via KL-divergence over all target samples:
\begin{equation}
\mathcal{L}_{\text{KD}} = \frac{1}{|\mathcal{D}_t|} \sum_{i \in \mathcal{D}_t} \sum_{k=1}^{K} p_{i,k}^{\text{FM},\text{L}} \log \frac{p_{i,k}^{\text{FM},\text{L}}}{p_{i,k}^{\text{SM},\text{L}}}
\label{eq:kd_loss}
\end{equation}
Unlike $\mathcal{L}_{\text{CE}}$, knowledge distillation operates on all samples since soft probability distributions from the calibrated FM provide more robust supervision than hard pseudo-labels.

\paragraph{Diversity Regularization.}
A common failure mode in SFDA is solution collapse, where the model assigns most samples to a few dominant classes. To prevent this, we maximize the entropy of the mean prediction over all target samples:
\begin{equation}
\mathcal{L}_{\text{Div}} = \sum_{k=1}^{K} \bar{p}_{k}^{\text{SM}} \log \bar{p}_{k}^{\text{SM}}, \quad \text{where} \quad \bar{p}_{k}^{\text{SM}} = \frac{1}{|\mathcal{D}_t|} \sum_{i \in \mathcal{D}_t} p_{i,k}^{\text{SM},\text{L}}
\label{eq:div_loss}
\end{equation}
This regularizer encourages balanced class predictions across the target domain, computed over all samples to accurately estimate the global distribution.

\subsection{Training Objectives}
\label{sec:objectives}
We summarize the overall optimization objectives for both branches during target adaptation.

\paragraph{Foundation Model.}
The FM is optimized solely through the mutual information calibration loss:
\begin{equation}
\mathcal{L}^{\text{FM}} = \mathcal{L}_{\text{MI}}
\label{eq:total_fm}
\end{equation}
where only the classifier $\mathcal{G}_{\text{FM}}$ is updated while the backbone $\mathcal{F}_{\text{FM}}$ remains frozen to preserve pre-trained representations.

\paragraph{Specialist Model.}
The SM is optimized through a joint objective combining pseudo-label supervision, knowledge distillation, and diversity regularization:
\begin{equation}
\mathcal{L}^{\text{SM}} = \mathcal{L}_{\text{CE}} + \lambda_{\text{kd}} \mathcal{L}_{\text{KD}} + \lambda_{\text{div}} \mathcal{L}_{\text{Div}}
\label{eq:total_sm}
\end{equation}
where $\lambda_{\text{kd}}$ and $\lambda_{\text{div}}$ are trade-off hyperparameters. The backbone $\mathcal{F}_{\text{SM}}$ is updated while the classifier $\mathcal{G}_{\text{SM}}$ remains frozen to preserve source-domain decision boundaries.

Following the calibrate-then-distill paradigm, both branches are optimized jointly in each iteration: the FM classifier is calibrated via $\mathcal{L}^{\text{FM}}$, and the SM backbone is adapted via $\mathcal{L}^{\text{SM}}$.
%%%%%%%%%%

\section{Experiment}
\subsection{Experimental Setup}
\subsubsection{Datasets and Settings.}
We evaluate FUSED on three widely-used EEG benchmarks spanning distinct paradigms: BCIC-IV-2a\cite{brunner2008bci}, FACED\cite{chen2023large}, and SSVEP\cite{wang2016benchmark}. Dataset details are provided in the \textit{supplementary file}.
We compare against state-of-the-art SFDA methods, with detailed descriptions in the \textit{supplementary file}. Source-only performance is included as a lower bound. We report mean accuracy across all cross-subject scenarios.

% 表格2
\begin{table*}[htbp]
\centering
\caption{Cross-subject classification accuracy (\%) of different methods on the FACED using the LOGO protocol. The best results are highlighted in \textbf{bold}, and the second-best results are \underline{underlined}.}
\label{tab:group_results}
% \small
\renewcommand{\arraystretch}{1.05} % 默认是 1，调小可以压缩行高
\setlength{\tabcolsep}{7.5pt} 
\begin{tabular}{lccccccccccc}
\toprule
\textbf{Method} & \textbf{G1} & \textbf{G2} & \textbf{G3} & \textbf{G4} & \textbf{G5} & \textbf{G6} & \textbf{G7} & \textbf{G8} & \textbf{G9} & \textbf{G10} & \textbf{Avg.} \\ 
\midrule
Source-only & 45.80 & 56.05 & 57.03 & 50.77 & 53.16 & 58.87 & 60.34 & 53.86 & 53.24 & 55.76 & 54.49 \\
\midrule
SHOT\cite{liang2020we}        & \underline{47.88} & 58.10 & 57.15 & 55.17 & \underline{57.29} & \underline{62.35} & 63.73 & 55.28 & 60.09 & 60.88 & 57.79 \\
AaD\cite{yang2022attracting}         & 45.28 & 60.90 & 58.34 & 55.72 & 56.11 & \underline{62.35} & 60.28 & \underline{58.48} & 60.22 & 58.76 & 57.64 \\
NRC\cite{yang2021exploiting}         & 45.21 & 59.84 & 58.27 & 54.85 & 55.28 & 60.72 & 59.34 & 56.20 & 59.82 & 58.04 & 56.76 \\
MAPU\cite{ragab2023source}        & 46.23 & 63.90 & 62.73 & 54.98 & 57.00 & 61.75 & \underline{65.31} & 55.33 & \underline{62.08} & \underline{61.04} & 59.03 \\
$\text{SF(DA)}^2$\cite{DBLP:conf/iclr/Hwang0SY24} & 46.40 & \underline{64.89} & \textbf{66.03} & 55.48 & 56.21 & 61.90 & 63.78 & 58.44 & \textbf{63.00} & 60.51 & \underline{59.66} \\
E-MAPU\cite{ragab2025evidentially}      & 46.32 & 62.79 & 59.40 & \textbf{56.33} & 55.10 & 60.95 & 65.06 & 57.78 & 61.93 & 59.88 & 58.55 \\
\midrule
\rowcolor{gray!20}
\textbf{FUSED (Ours)} & \textbf{49.98} & \textbf{67.36} & \underline{65.00} & \underline{56.12} & \textbf{59.65} & \textbf{65.07} & \textbf{66.38} & \textbf{59.57} & 61.32 & \textbf{61.21} & \textbf{61.16} \\
\bottomrule
\end{tabular}
\end{table*}

% 表格3
\begin{table*}[htbp]
\centering
\caption{Cross-subject classification accuracy (\%) of different methods on the SSVEP Benchmark using the LOSO protocol. The best results are highlighted in \textbf{bold}, and the second-best results are \underline{underlined}.}
\label{tab:ssvep_table}
% \small
\renewcommand{\arraystretch}{1.05} % 默认是 1，调小可以压缩行高
\setlength{\tabcolsep}{7.5pt} 
\begin{tabular}{lccccccccccc}
\toprule
\textbf{Method} & \textbf{S1} & \textbf{S2} & \textbf{S3} & \textbf{S4} & \textbf{S5} & \textbf{S6} & \textbf{S7} & \textbf{S8} & \textbf{S9} & \textbf{S10} & \textbf{Avg.} \\ 
\midrule
Source-only & 80.83 & 70.00 & 77.71 & 78.96 & 82.29 & 85.00 & 85.83 & 86.88 & 63.75 & 82.50 & 74.28 \\
\midrule
SHOT\cite{liang2020we}        & 82.50 & 80.83 & 81.04 & 79.58 & 82.08 & 85.13 & 87.50 & 85.62 & 64.38 & 82.92 & 77.01 \\
AaD\cite{yang2022attracting}         & 81.38 & 70.44 & 80.26 & 81.35 & 83.24 & 85.71 & 86.15 & 87.38 & 68.54 & 83.12 & 76.78 \\
NRC\cite{yang2021exploiting}         & 80.91 & 70.78 & 79.34 & 81.82 & 84.01 & 86.77 & 86.42 & 88.15 & 64.58 & 84.97 & 77.42 \\
MAPU\cite{ragab2023source}        & \underline{83.52} & 80.95 & 82.12 & 85.21 & \underline{89.49} & 86.16 & 86.39 & 88.26 & 70.06 & 84.32 & 77.85 \\
$\text{SF(DA)}^2$\cite{DBLP:conf/iclr/Hwang0SY24} & 83.29 & \underline{81.63} & \underline{83.50} & \underline{88.95} & 89.47 & \underline{88.68} & \textbf{91.35} & \underline{90.34} & \underline{70.80} & \underline{87.42} & \underline{78.33} \\
E-MAPU\cite{ragab2025evidentially}      & 82.88 & 80.24 & 79.28 & 85.49 & 88.20 & 87.41 & 88.79 & 87.17 & 70.01 & 83.00 & 77.80 \\
\midrule
\rowcolor{gray!20}
\textbf{FUSED (Ours)} & \textbf{83.96} & \textbf{81.67} & \textbf{85.21} & \textbf{92.29} & \textbf{90.00} & \textbf{89.17} & \underline{90.00} & \textbf{90.62} & \textbf{72.08} & \textbf{89.92} & \textbf{80.95} \\
\bottomrule
\end{tabular}
\end{table*}

\subsubsection{Evaluation Protocol.}  
We adopt dataset-specific cross-subject validation schemes. For BCIC-IV-2a and SSVEP Benchmark, we employ \textit{Leave-One-Subject-Out} (LOSO), where each subject serves as the target domain $\mathcal{D}_t$ while remaining subjects form $\mathcal{D}_s$. For the larger-scale FACED dataset (123 subjects), we use \textit{Leave-One-Group-Out} (LOGO), partitioning subjects into groups of ten with one group as $\mathcal{D}_t$ per fold. All baselines are re-implemented under identical configurations for fair comparison.

\subsubsection{Network Architectures.}
The FM branch employs CbraMod~\cite{DBLP:conf/iclr/WangZLZJLL025} as the backbone, followed by a projection layer (Linear-ELU-BatchNorm) mapping to a $D^{\text{FM}}=200$ dimensional space. The SM branch adopts EEGNet~\cite{lawhern2018eegnet} with temporal, spatial depthwise, and fusion convolutions, producing $D^{\text{SM}}=128$ dimensional features. Both branches use a single linear layer as the classifier.

\subsubsection{Hyperparameters and Optimization.}
We train for 50 epochs with batch sizes of 32 (BCIC-IV-2a, SSVEP) and 64 (FACED). We use Adam optimizer with initial learning rate $1 \times 10^{-4}$ and exponential decay (power 0.75). Method-specific hyperparameters are: EMA momentum $\mu = 0.9$, margin threshold $\eta = 0.6$, and loss weights $\lambda_{\text{kd}} = \lambda_{\text{div}} = 1.0$. All experiments are conducted in PyTorch on an NVIDIA RTX 4090 GPU.

\subsection{Comparative Results}
Tables~\ref{tab:loso_results},~\ref{tab:group_results}, and~\ref{tab:ssvep_table} present the cross-subject classification performance on three EEG tasks: motor imagery, emotion recognition, and SSVEP. For SSVEP, we report results for the first ten subjects due to space constraints (complete results in the \textit{supplementary file}), whereas the average accuracy (Avg.) here is computed across the entire cohort of 35 subjects.

\paragraph{Overall Performance}
FUSED achieves state-of-the-art performance across all three benchmarks, outperforming the runner-up by {1.07\%} on BCIC-IV-2a, {1.50\%} on FACED, and {2.62\%} on SSVEP. All SFDA methods substantially outperform the Source-only baseline, confirming the necessity of target-domain adaptation for mitigating inter-subject variability in EEG decoding.

\paragraph{Baseline Analysis.}
A detailed comparison reveals the specific challenges inherent in EEG-based SFDA. SHOT relies on pseudo-labels from all target samples without quality filtering, making it susceptible to label noise and unstable adaptation. AaD and NRC exploit local geometric structures in feature space, but pure geometric alignment proves insufficient under the severe non-stationarity of EEG signals. MAPU and E-MAPU incorporate temporal modeling, yielding improved performance on FACED and SSVEP where temporal dynamics are critical. SF(DA)$^2$ achieves consistent runner-up performance through implicit feature augmentation, yet remains limited by its single-model architecture.

% 表格4 消融
\begin{table*}[t]
\centering
\caption{Ablation study of the proposed components on three EEG datasets. The results represent the average classification accuracy (\%). ``\cmark'' indicates the inclusion of a component, while ``\xmark'' denotes its removal from the full framework.}
\label{tab:ablation_study}
% \small
\renewcommand{\arraystretch}{1.1} % 默认是 1，调小可以压缩行高
\setlength{\tabcolsep}{8.5pt} 
\begin{tabular}{cccccccc}
\toprule
\textbf{Consensus Mask} & \textbf{MI loss} & \textbf{KL Loss} & \textbf{CE Loss} & \textbf{Div Loss} & \textbf{BCIC-IV-2a} & \textbf{FACED} & \textbf{SSVEP} \\ 
\midrule
\xmark & \cmark & \cmark & \cmark & \cmark & 67.35 & 59.79 & 79.91 \\
\xmark & \cmark & \cmark & \xmark         & \cmark & 65.52 & 58.64 & 78.25 \\
\cmark & \xmark         & \cmark & \cmark & \cmark & 67.40 & 59.73 & 78.70 \\
\cmark & \cmark & \xmark         & \cmark & \cmark & 67.11 & 59.43 & 78.77 \\
\cmark & \cmark & \cmark & \cmark & \xmark        & 67.51 & 59.78 & 80.54 \\
\midrule
\rowcolor{gray!20}
\cmark & \cmark & \cmark & \cmark & \cmark & \textbf{68.77} & \textbf{61.16} & \textbf{80.95} \\
\bottomrule
\end{tabular}
\end{table*}

\paragraph{Superiority of FUSED.}
Our method overcomes the inherent ceiling of single-model SFDA by leveraging FM-SM synergy. The FM's robust generalization priors, derived from large-scale pre-training, compensate for representation deficiencies when task-specific data is limited. Meanwhile, the dual-view pseudo-label refinement, combining consensus filtering with cross-branch arbitration, effectively suppresses error accumulation across all three paradigms. This calibrate-then-distill strategy enables versatile adaptation to diverse EEG tasks while maintaining classification precision.

\subsection{Ablation Study}
We conduct ablation studies to validate each component's contribution, examining individual loss functions and consensus filtering, dual-branch versus single-branch architectures, and two-stage pseudo-label refinement versus single-view alternatives.

\subsubsection{Component Analysis}
Table~\ref{tab:ablation_study} presents the contribution of each component across three datasets. The full framework consistently achieves the highest accuracy, demonstrating synergistic effects among modules.
Removing both the Consensus Mask and CE Loss causes the most significant degradation (68.77\% → 65.52\% on BCIC-IV-2a), confirming that dual-view consensus filtering is fundamental for reliable self-training. Without it, noisy pseudo-labels trigger error accumulation. Removing MI Loss leads to notable drops across all tasks (\emph{e.g.}, 2.25\% on SSVEP), validating that FM calibration is essential for generating high-quality teacher signals. Similarly, removing KL Loss degrades performance, confirming the importance of transferring FM knowledge to the SM. Finally, removing Div Loss yields marginal but consistent decline, indicating its role as a safeguard against solution collapse.

\subsubsection{Dual-Branch Architecture}
Table~\ref{tab:dual_branch} compares single-branch baselines against our dual-branch framework. Both variants use identical optimization for fair comparison.
Single-branch models exhibit complementary strengths: the FM generalizes better on FACED, while the SM excels on BCIC-IV-2a. Our dual-branch framework surpasses both by effectively combining their advantages. Notably, despite managing two branches, strategic freezing of the FM backbone keeps computational overhead competitive—epoch time (0.41s on BCIC-IV-2a) remains lower than full FM adaptation (1.23s), providing an favorable accuracy-efficiency trade-off.

\begin{table}[t]
\centering
\caption{Adaptation performance and computational overhead analysis across single-branch baselines and the proposed dual-branch framework.}
\label{tab:dual_branch}
% \small
\renewcommand{\arraystretch}{1.2} % 默认是 1，调小可以压缩行高
\begin{tabular}{lcccc}
\toprule
\multirow{2}{*}{Variant} & \multicolumn{2}{c}{BCIC-IV-2a} & \multicolumn{2}{c}{FACED} \\
\cmidrule(lr){2-3} \cmidrule(lr){4-5}
                         & Acc(\%)   & Epoch time(s) & Acc(\%)   & Epoch time(s) \\
\midrule
FM branch                & 60.21     & 1.23          & 57.89     & 3.00 \\
SM branch                & 63.32     & 0.13          & 55.43     & 0.79 \\
\midrule
\rowcolor{gray!20}
Dual branch              & \textbf{68.77}     & 0.41          & \textbf{61.16}     & 1.71 \\
\bottomrule
\end{tabular}
\end{table}

\subsubsection{Pseudo-Label Refinement Strategy}
Table~\ref{tab:pseudo_label_results} evaluates the two-stage refinement mechanism against four single-view variants: $p^{\text{FM},\text{P}}$, $p^{\text{FM},\text{L}}$, $p^{\text{SM},\text{P}}$, and $p^{\text{SM},\text{L}}$, representing pseudo-labels from individual views. All other components remain active to isolate the refinement contribution.
Relying on any single view leads to suboptimal adaptation. On FACED, SM-only views show significant drops, indicating that task-specific models lack robustness for clean pseudo-labeling in early adaptation. Interestingly, the relative effectiveness of FM versus SM views varies across datasets—SM views perform competitively on BCIC-IV-2a but underperform on FACED. This inconsistency underscores the risk of single-view dependency and validates our cross-branch arbitration strategy.

\begin{table}[t]
\centering
\caption{Ablation study of different pseudo-labels strategies}
\label{tab:pseudo_label_results}
% \small
\renewcommand{\arraystretch}{1.1} % 默认是 1，调小可以压缩行高
\setlength{\tabcolsep}{16pt} 
\begin{tabular}{lcc}
\toprule
Variants & BCIC-IV-2a & FACED \\
\midrule
\textit{\( p^{FM,P} \)}      & 67.09                   & 59.43 \\
\textit{\( p^{FM,L} \)}      & 66.54                   & 60.21 \\
\textit{\( p^{SM,P} \)}       & 67.65                   & 58.20 \\
\textit{\( p^{SM,L} \)}       & 67.33                   & 58.74 \\
\rowcolor{gray!20}
\midrule
\textbf{FUSED (Ours)}               & \textbf{68.77}                   & \textbf{61.16} \\
\bottomrule
\end{tabular}
\end{table}

% compare different backbones
\begin{figure}[t]
    \centering
    \includegraphics[width=0.85\columnwidth]{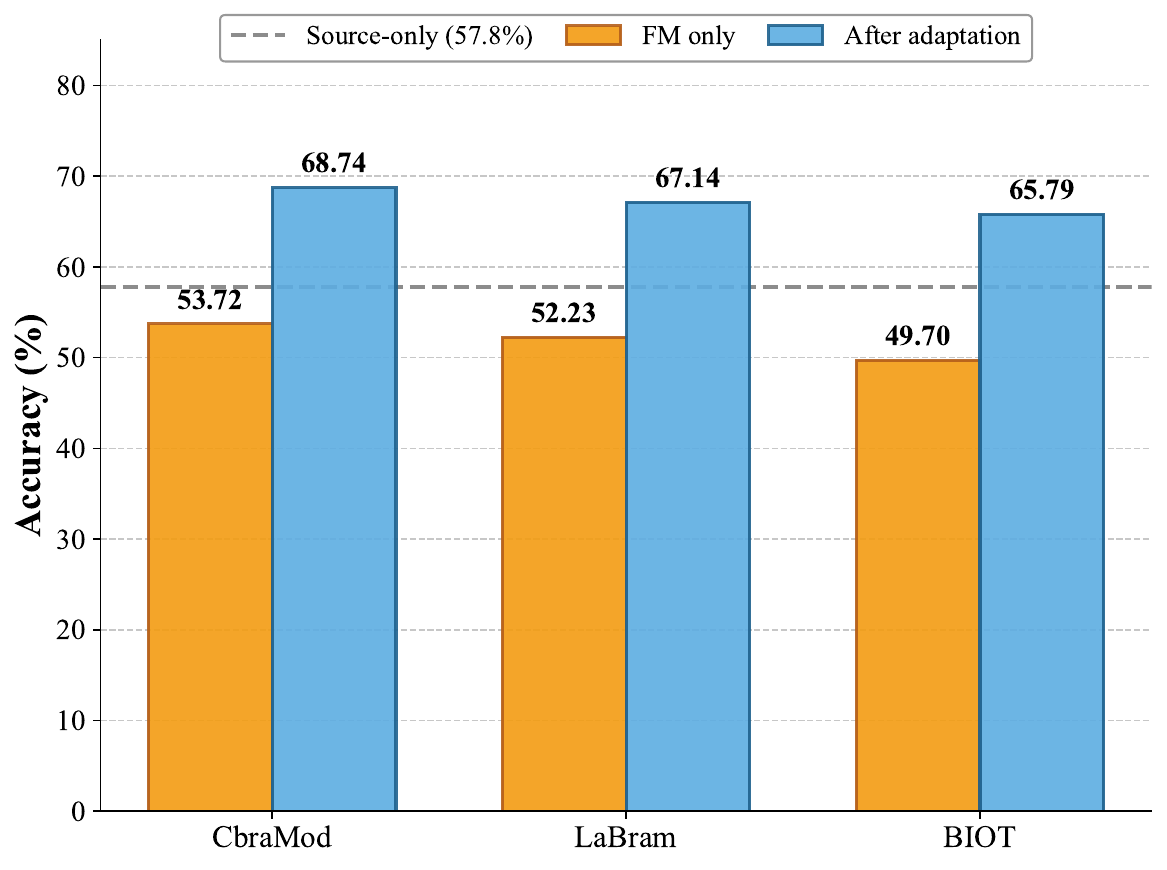}
    \caption{Compare different foundation model backbones on BCIC-IV-2a dataset}
    \label{fig:backbone_compare}
\end{figure}

% 参数调节
\begin{figure*}
    \centering
    \begin{minipage}{0.24\linewidth}
        \centering
        \includegraphics[width=\linewidth]{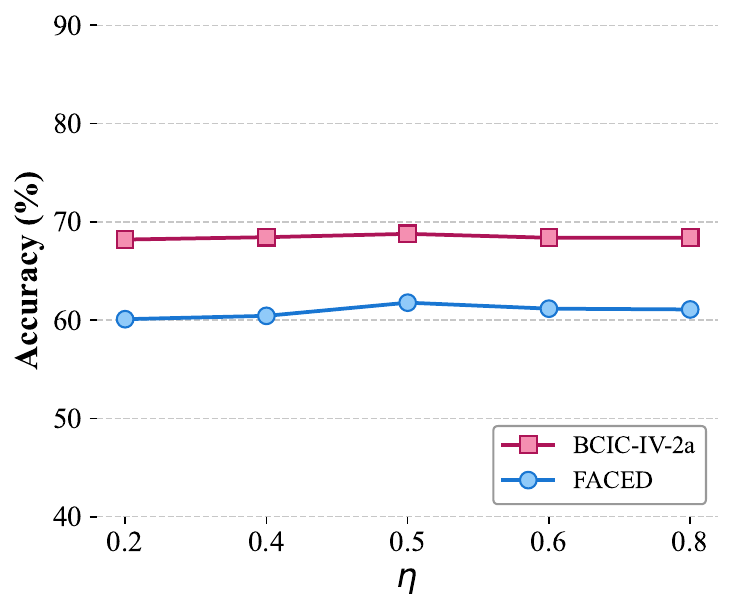}
        \centerline{\footnotesize (a) Margin Threshold $\eta$}
    \end{minipage}
    \hfill
    \begin{minipage}{0.24\linewidth}
        \centering
        \includegraphics[width=\linewidth]{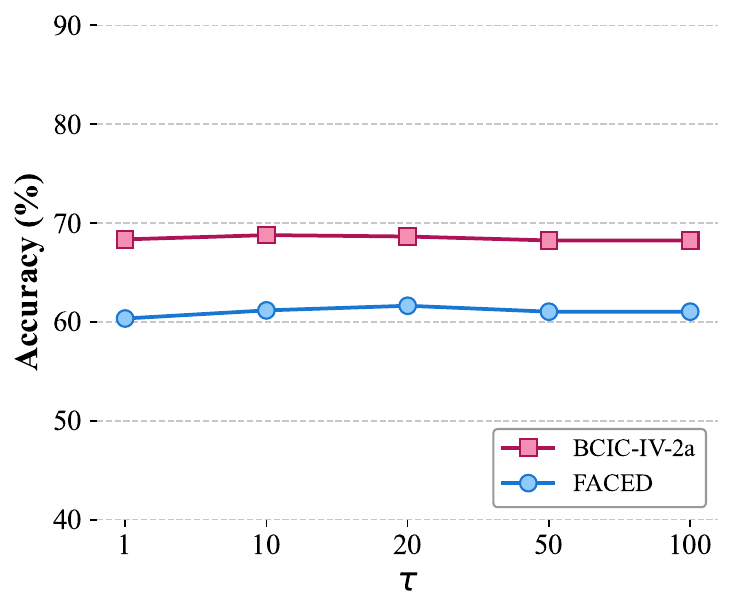}
        \centerline{\footnotesize (b) Scaling Parameter $\tau$}
    \end{minipage}
    \hfill
    \begin{minipage}{0.24\linewidth}
        \centering
        \includegraphics[width=\linewidth]{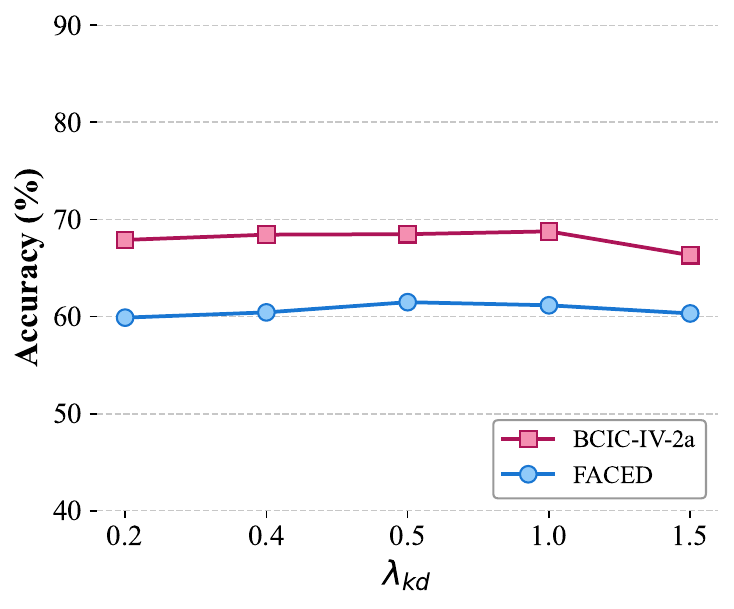}
        \centerline{\footnotesize (c) KL Weight $\lambda_{kd}$}
    \end{minipage}
    \hfill
    \begin{minipage}{0.24\linewidth}
        \centering
        \includegraphics[width=\linewidth]{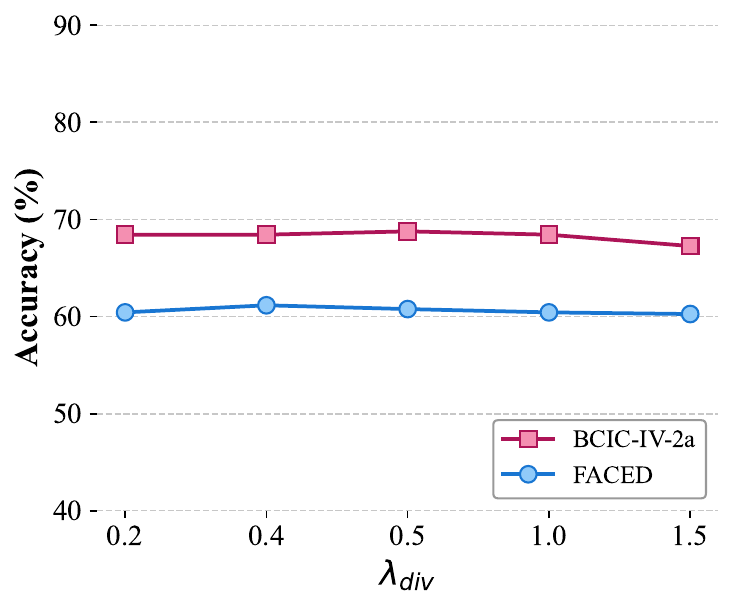}
        \centerline{\footnotesize (d) Div Weight $\lambda_{div}$}
    \end{minipage}

    \caption{Parameter sensitivity analysis on BCIC-IV-2a and FACED datasets. The classification accuracy (\%) is reported under varying (a) margin threshold $\eta$, (b) scaling parameter $\tau$, (c) knowledge distillation weight $\lambda_{kd}$, and (d) diversity loss weight $\lambda_{div}$.}
    \label{fig:parameter_sensitivity}
\end{figure*}

\begin{figure}[t]
    \centering
    % --- 左图：Before Adaptation ---
    \begin{minipage}{0.49\linewidth}
        \centering
        \includegraphics[width=\linewidth]{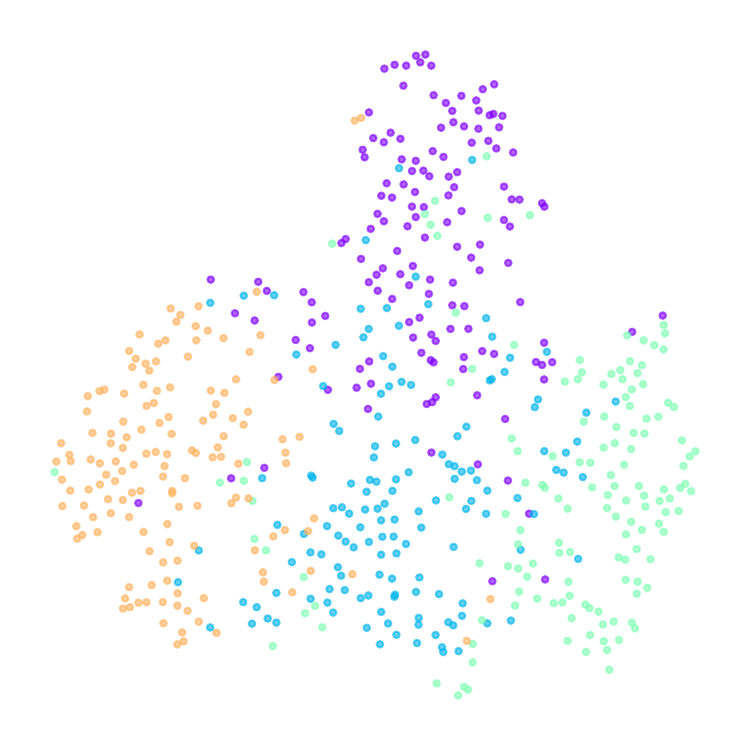}
        \centerline{\footnotesize (a)}
    \end{minipage}
    \hfill
    % --- 右图：After Adaptation ---
    \begin{minipage}{0.49\linewidth}
        \centering
        \includegraphics[width=\linewidth]{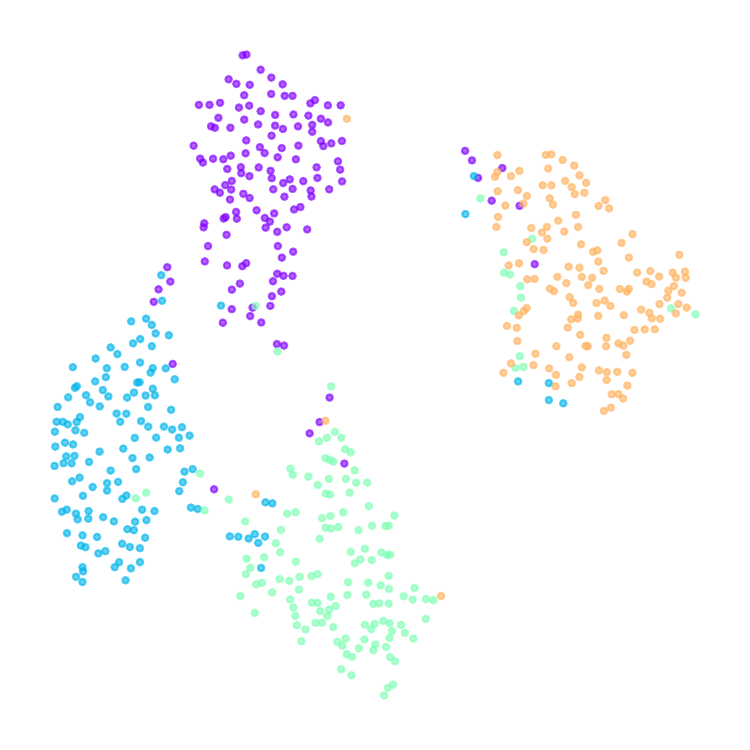}
        \centerline{\footnotesize (b)}
    \end{minipage}

    \caption{t-SNE visualization of extracted features on BCIC-IV-2a. Left illustrates the features from the source pre-trained model without adaptation. Right depicts the features after adaptation using our proposed framework. Each color represents a distinct category.}
    \label{fig:tsne_visualization}
\end{figure}

\subsection{Model Analysis}
We present further analyses to validate the robustness and generalizability of FUSED, examining compatibility with different FM backbones, sensitivity to hyperparameter settings, and feature visualization before and after adaptation.

\subsubsection{FM Backbone Compatibility}
We evaluate FUSED with three mainstream EEG foundation models: CbraMod~\cite{DBLP:conf/iclr/WangZLZJLL025}, LaBraM~\cite{DBLP:conf/iclr/JiangZL24}, and BIOT~\cite{yang2023biot}. Figure~\ref{fig:backbone_compare} compares two scenarios on BCIC-IV-2a: (i) \textit{FM-only}, where the FM is fine-tuned on source data and directly evaluated on the target domain (analogous to source-only), and (ii) \textit{After adaptation}, representing performance after applying FUSED.
All three backbones exhibit substantial performance gains after adaptation, with improvements of 15.02\%, 14.91\%, and 16.09\% for CbraMod, LaBraM, and BIOT, respectively. This consistent trend across diverse architectures demonstrates that our framework's effectiveness stems from the proposed adaptation mechanism rather than the inherent capability of any specific FM backbone.

\subsubsection{Parameter Sensitivity Analysis}
Figure~\ref{fig:parameter_sensitivity} illustrates the impact of four key hyperparameters on BCIC-IV-2a and FACED. Our framework exhibits consistent stability across a wide range of settings. The margin threshold $\eta$ performs optimally within $[0.4, 0.6]$, where consensus filtering effectively removes low-quality pseudo-labels. The temperature $\tau$ for prototype prediction shows high robustness, with best results at $\tau \in \{10, 20\}$. The loss weights $\lambda_{\text{kd}}$ and $\lambda_{\text{div}}$ maintain stable accuracy near 1.0, with only marginal decline at extreme values. These consistent trends across both datasets confirm that FUSED achieves reliable adaptation without exhaustive hyperparameter tuning.

\subsubsection{Visualization Analysis}
We visualize feature distributions using t-SNE to qualitatively assess adaptation effectiveness. As shown in Figure~\ref{fig:tsne_visualization}, features from the source pre-trained model (left) exhibit significant class overlap and scattered distributions, reflecting the domain discrepancy that hinders clear decision boundaries.
After adaptation (right), target features demonstrate markedly improved separation—samples aggregate into cohesive class-specific clusters with clear inter-class margins. This visual evidence confirms that our dual-view refinement and knowledge transfer strategies effectively mitigate domain shift, enabling the SM to learn discriminative, task-specific representations.

\section{Conclusion}
In this paper, we presented FUSED, a novel framework for source-free domain adaptation in EEG decoding that addresses the challenges of inter-subject variability and signal non-stationarity. Unlike existing single-model approaches that are susceptible to noisy pseudo-labels and error accumulation, FUSED introduces a dual-branch co-adaptation mechanism that synergistically combines the robust generalization of Foundation Models with the task-specific precision of Specialist Models. Through Consensus Filtering and Two-Stage Pseudo-Label Refinement, our framework effectively suppresses error propagation, while the calibrate-then-distill strategy enables principled knowledge transfer without source data access. To our knowledge, FUSED is the first framework to leverage large-scale EEG foundation models within the source-free adaptation paradigm. Extensive experiments across three diverse EEG tasks demonstrate consistent state-of-the-art performance, establishing FUSED as a robust and privacy-preserving solution for real-world BCI applications.
Some promising directions remain for future work. First, extending FUSED to open-set scenarios where target domains contain unknown classes would enhance its practical applicability. Additionally, exploring cross-dataset transfer—adapting models trained on one EEG paradigm to entirely different tasks—could further validate the generalizability of foundation-guided adaptation.

% \clearpage  % 或使用 \usepackage{placeins} 后的 \FloatBarrier

% \section*{Acknowledgments}
% This should be a simple paragraph before the References to thank those individuals and institutions who have supported your work on this article.

\bibliographystyle{IEEEtran}
\bibliography{refs}

% {\appendix
% \maketitle
\appendix
\section{Description of Datasets}
\subsection{BCIC-IV-2a Dataset}
We use the well-known BCI Competition 2a dataset\cite{brunner2008bci} for motor imagery classification. It contains EEG recordings from 9 subjects performing motor imagery of four classes(left hand, right hand, both feet, tongue). Raw EEG signals were recorded at 22 channels and 250 Hz sampling rate. Each subject performed two sessions on different days, with a total of 288 trails per session. The collected data underwent the following preprocessing which included applying  a bandpass filter with a frequency range of 0.3-50 Hz and resampling to 200 Hz. For each trail, we extracted a 4-second window from 2s to 6s after the cue onset. This results in a final input shape of $\mathbb{R}^{22 \times 4 \times 200}$ for each sample.

\subsection{FACED Dataset}
This dataset\cite{chen2023large} is specifically designed for fine-grained emotion recognition which comprises data from 123 subjects, involving the use of 28 video clips as stimuli to elicit negative, positive and neutral emotions. Each video clip has an average duration of approximately 67s. These video clips are associated with a range of emotion items, including anger, disgust, fear, sadness, amusement, joy, inspiration, tenderness, arousal, valence, familiarity, and liking. After viewing each video clip, subjects provided self-report emotional scores for these emotion items. The raw EEG data was recorded using a 32-channel EEG system at a sampling rate of 250 Hz. The collected data underwent preprocessing, which included applying a bandpass filter with a frequency range of 0.05-47 Hz and resampling to 200 Hz. In addition, independent component analysis(ICA) was employed to effectively remove artifacts from the EEG data. Each 30-second trial was further segmented into three non-overlapping 10-second windows via a sliding window approach. This results in a final input shape of $\mathbb{R}^{32 \times 10 \times 200}$ for each sample. 

\subsection{SSVEP Dataset}
The TSU dataset\cite{wang2016benchmark} is a benchmark dataset for BCI based on steady-state visual evoked potentials. This dataset collects recordings from 35 healthy subjects using a brain-computer interface(BCI) speller with 40 characters that flashed at different frequencies(8-15.8 Hz with 0.2 Hz intervals), corresponding to 40 classes. The raw EEG data was acquired using a 64-channel system at a sampling rate of 250 Hz,  with each subject completing 6 blocks of 40 trials each. The collected data underwent the following preprocessing which included 9 effective channel selection(Indices: 47, 53-57, and 60-62), resampling to 200 Hz. For each 6-second trial, we extracted 4-second epoch (from 1s to 5s post-stimulus) to eliminate the transient evoked potentials at the stimulus onset and offset. The 4-second resampled epochs were further segmented into two non-overlapping 2-second windows via a sliding window approach. This results in a final input shape of $\mathbb{R}^{9 \times 2 \times 200}$ for each sample. 

\begin{table}[htbp]
\centering
\caption{Details of the adopted datasets (C: \#channels, K: \#classes, L: sample length).}
\label{tab:dataset_details}
\begin{tabular}{lcccccc}
\toprule
Dataset & C & K & L & \#Train samples & \#Test samples \\
\midrule
BCIC-IV-2a & 22 & 4 & 800 & 4608 & 576 \\
FACED & 32 & 9 & 2000 & 9240 & 1092 \\
SSVEP & 9 & 40 & 400 & 16320 & 480 \\
\bottomrule
\end{tabular}
\end{table}

\begin{table*}[t]
\centering
\caption{Cross-subject classification accuracy (\%) of different methods on the SSVEP Benchmark using the LOSO protocol (all 35 subjects). The best results are highlighted in \textbf{bold}, and the second-best results are \underline{underlined}.}
\label{tab:ssvep_full}
\renewcommand{\arraystretch}{1.05}
\setlength{\tabcolsep}{5pt}
\begin{tabular}{lcccccccc}
\toprule
\textbf{Subject} & \textbf{Source-only} & \textbf{SHOT} & \textbf{AaD} & \textbf{NRC} & \textbf{MAPU} & \textbf{SF(DA)$^2$} & \textbf{E-MAPU} & \textbf{FUSED (Ours)} \\
\midrule
S1 & 80.83 & 82.50 & 81.38 & 80.91 & 83.52 & 83.29 & 82.88 & \textbf{83.96} \\
S2 & 70.00 & 80.83 & 70.44 & 70.78 & 80.95 & 81.63 & 80.24 & \textbf{81.67} \\
S3 & 77.71 & 81.04 & 80.26 & 79.34 & 82.12 & 83.50 & 79.28 & \textbf{85.21} \\
S4 & 78.96 & 79.58 & 81.35 & 81.82 & 85.21 & 88.95 & 85.49 & \textbf{92.29} \\
S5 & 82.29 & 82.08 & 83.24 & 84.01 & 89.49 & 89.47 & 88.20 & \textbf{90.00} \\
S6 & 85.00 & 85.13 & 85.71 & 86.77 & 86.16 & 88.68 & 87.41 & \textbf{89.17} \\
S7 & 85.83 & 87.50 & 86.15 & 86.42 & 86.39 & \textbf{91.35} & 88.79 & \underline{90.00} \\
S8 & 86.88 & 85.62 & 87.38 & 88.15 & 88.26 & 90.34 & 87.71 & \textbf{90.62} \\
S9 & 63.75 & 64.38 & 68.54 & 64.58 & 70.06 & 70.80 & 70.01 & \textbf{72.08} \\
S10 & 82.50 & 82.90 & 83.12 & 84.97 & 84.32 & 87.42 & 83.00 & \textbf{89.92} \\
S11 & 55.28 & 58.24 & 57.36 & 57.07 & 57.36 & 60.42 & 59.48 & \textbf{61.88} \\
S12 & 68.17 & 69.58 & 69.22 & 70.22 & 70.92 & 75.25 & 74.30 & \textbf{77.08} \\
S13 & 73.38 & 79.04 & 74.92 & 76.08 & 79.14 & 79.66 & 79.50 & \textbf{84.17} \\
S14 & 44.72 & 48.75 & 46.07 & 46.92 & 47.67 & 49.90 & 48.62 & \textbf{49.58} \\
S15 & 84.67 & 88.31 & 86.30 & 87.35 & 88.30 & 90.74 & 89.88 & \textbf{91.04} \\
S16 & 49.10 & 50.47 & 53.11 & 53.31 & 50.25 & 51.21 & 50.17 & \textbf{56.46} \\
S17 & 83.93 & 85.92 & 88.42 & 88.90 & 84.29 & 84.68 & 84.46 & \textbf{91.25} \\
S18 & 80.69 & 83.07 & 83.19 & 83.52 & 83.27 & 74.09 & 72.44 & \textbf{75.21} \\
S19 & 50.04 & 53.15 & 53.28 & 54.37 & 53.61 & 55.38 & 54.35 & \textbf{58.24} \\
S20 & 80.77 & 83.79 & 83.97 & 84.33 & 84.77 & 84.52 & 83.27 & \textbf{86.88} \\
S21 & 82.19 & 85.10 & 85.04 & 85.68 & 85.00 & 85.68 & 85.34 & \textbf{87.71} \\
S22 & 85.31 & 89.33 & 88.15 & 89.17 & 88.73 & 90.46 & 89.85 & \textbf{92.71} \\
S23 & 60.43 & 64.73 & 63.39 & 64.25 & 64.10 & 64.33 & 64.20 & \textbf{68.54} \\
S24 & 82.76 & 84.91 & 85.73 & 85.34 & 85.87 & 85.97 & 86.01 & \textbf{87.50} \\
S25 & 78.17 & 80.47 & 82.00 & 81.15 & 81.21 & 80.40 & 83.31 & \textbf{85.99} \\
S26 & 85.93 & 88.02 & 88.74 & 88.31 & 88.47 & 90.72 & 88.72 & \textbf{95.00} \\
S27 & 87.06 & 90.67 & 90.17 & 89.26 & 90.20 & 90.88 & 91.24 & \textbf{95.62} \\
S28 & 81.70 & 83.90 & 84.08 & 85.48 & 83.88 & 84.46 & 81.08 & \textbf{87.08} \\
S29 & 42.95 & 44.88 & 46.13 & 48.27 & 44.12 & 41.65 & 45.21 & \textbf{47.29} \\
S30 & 70.42 & 72.13 & 70.88 & 71.18 & 72.32 & \textbf{73.51} & 72.82 & \underline{72.50} \\
S31 & 90.21 & 92.20 & 92.77 & 93.40 & 93.98 & 94.17 & 93.96 & \textbf{99.38} \\
S32 & 93.86 & 95.84 & 97.58 & 97.38 & 97.50 & 97.32 & 97.38 & \textbf{98.75} \\
S33 & 25.58 & 25.63 & 31.74 & \textbf{34.15} & 28.18 & 23.51 & 28.47 & \underline{26.46} \\
S34 & 85.87 & 91.32 & 93.11 & 92.89 & 91.55 & 87.18 & 92.78 & \textbf{95.34} \\
S35 & 82.86 & 93.64 & 90.28 & 93.97 & 93.58 & 90.03 & 93.15 & \textbf{96.67} \\
\midrule
\textbf{Avg.} & 74.28 & 77.01 & 76.78 & 77.42 & 77.85 & \underline{78.33} & 77.80 & \textbf{80.95} \\
\bottomrule
\end{tabular}
\end{table*}

\section{Description of Baselines}
To comprehensively evaluate the performance of FUSED, we compare it against six state-of-the-art Source-Free Domain Adaptation (SFDA) methods. These baselines represent diverse technical trajectories in the field:
\begin{itemize}
    \item SHOT\cite{liang2020we} : A pioneering SFDA framework that freezes the source classifier and adapts the feature extractor by maximizing information maximization loss with self-supervised pseudo-labels by computing the proximity of target features to class centroids in the latent space.
    
    \item AaD\cite{yang2022attracting} : This method facilitates adaptation by mining the local geometric topology of the feature space. It employs an ``attraction'' mechanism to align each sample with its K-nearest neighbors via KL-divergence, while a ``dispersal'' term prevents the model from converging to a degenerate solution.
    
    \item NRC\cite{yang2021exploiting} : Building upon neighborhood consistency, NRC introduces a more robust pseudo-labeling strategy by considering second-order neighbor affinities (i.e., neighbors of neighbors). It promotes label consistency within high-density regions to mitigate the impact of initial predictive noise.
    
    \item MAPU\cite{ragab2023source}: A task-agnostic SFDA method specifically engineered for time-series data. It introduces a temporal imputation pretext task where the model is required to recover randomly masked segments in the embedding space, thereby capturing intrinsic temporal dynamics to guide the adaptation.

    \item SF(DA)$^2$\cite{DBLP:conf/iclr/Hwang0SY24}: This approach leverages an augmentation graph in the latent space to guide spectral neighborhood clustering. By synergizing feature disentanglement with an implicit augmentation strategy, it effectively captures class-specific heuristics through the simulation of unbounded virtual samples, achieving robust boundary stabilization without increased computational costs.
    
    \item E-MAPU\cite{ragab2025evidentially}  : An extension of MAPU that incorporates Evidential Deep Learning (EDL) to quantify predictive uncertainty. It utilizes evidential entropy to identify out-of-support target samples and steer them toward the source domain's support, addressing the overconfidence issue common in standard softmax outputs.
    
\end{itemize}

\section{Detailed Results on SSVEP}
As presented in Table~\ref{tab:ssvep_full}, which provides comprehensive results across 35 subjects, our method consistently outperforms all six state-of-the-art baselines on the vast majority of subjects.This significant improvement validates the efficacy of our dual-branch co-adaptation mechanism, which synergistically combines the robust generalization of Foundation Models with the task-specific precision of Specialist Models, hence substantially enhancing both decoding accuracy and cross-subject robustness.

\clearpage  % 或使用 \usepackage{placeins} 后的 \FloatBarrier


\end{document}